%% file: Proc_LTD18_MPiat.tex
\pdfoutput=1
\documentclass[final]{svjour3}
\usepackage{graphicx}
\usepackage{rotating}
\usepackage{amssymb}
\usepackage{mathptmx}
\usepackage[numbers]{natbib}
\makeatletter
\journalname{Journal of Low Temperature Physics}

\usepackage{multirow}

\bibpunct{}{}{,}{n}{}{,}

\begin{document}

\newcommand{\hdblarrow}{H\makebox[0.9ex][l]{$\downdownarrows$}-}
\title{QUBIC: using NbSi TESs with a bolometric interferometer to characterize the polarisation of the CMB}

\input{Authors.tex}

\maketitle

\begin{abstract}

QUBIC (Q \& U Bolometric Interferometer for Cosmology) is an international ground-based experiment dedicated in the measurement of the polarized fluctuations of the Cosmic Microwave Background (CMB). It is based on bolometric interferometry, an original detection technique which combine the immunity to systematic effects of an interferometer with the sensitivity of low temperature incoherent detectors. QUBIC will be deployed in Argentina, at the Alto Chorrillos mountain site near San Antonio de los Cobres, in the Salta province.

The QUBIC detection chain consists in 2048 NbSi Transition Edge Sensors (TESs) cooled to 350mK.The voltage-biased TESs are read out with Time Domain Multiplexing based on Superconducting QUantum Interference Devices (SQUIDs) at 1 K and a novel SiGe Application-Specific Integrated Circuit (ASIC) at 60 K allowing to reach an unprecedented multiplexing (MUX) factor equal to 128. 

The QUBIC experiment is currently being characterized in the lab with a reduced number of detectors before upgrading to the full instrument. I will present the last results of this characterization phase with a focus on the detectors and readout system.


\end{abstract}

\section{Introduction}
\input{Intro.tex}

\section{The QUBIC instrument}
\label{sec:instru}
\input{Instru.tex}

\section{The QUBIC detection chain}
\label{sec:detchain}
\input{Det_chain.tex}

\section{Characterisation of the QUBIC detection chain}
\label{sec:charact}
\input{Charact.tex}

\section{Conclusions}
\input{Ccl.tex}

\begin{acknowledgements}
QUBIC is funded by the following agencies. France: ANR (Agence Nationale de la Recherche) 2012 and 2014, DIM-ACAV (Domaine dÕInteret MajeurÑAstronomie et Conditions dÕApparition de la Vie), CNRS/IN2P3 (Centre national de la recherche scientifique/Institut national de physique nuclŽaire et de physique des particules), CNRS/INSU (Centre national de la recherche scientifique/Institut national de sciences de lÕunivers), UnivEarthS Labex program at Sorbonne Paris Cit\'e (ANR-10-LABX-0023 and ANR-11-IDEX-0005-02). Italy: CNR/PNRA (Consiglio Nazionale delle Ricerche/Programma Nazionale Ricerche in Antartide) until 2016, INFN (Istituto Nazionale di Fisica Nucleare) since 2017. Argentina: Secretar'a de Gobierno de Ciencia, Tecnolog'a e Innovaci—n Productiva, Comisi—n Nacional de Energ'a At—mica, Consejo Nacional de Investigaciones Cient'ficas y TŽcnicas. UK: the University of Manchester team acknowledges the support of STFC (Science and Technology Facilities Council) grant ST/L000768/1. Ireland: James Murphy and David Burke acknowledge postgraduate scholarships from the Irish Research Council. Duc Hoang Thuong acknowledges the Vietnamese government for funding his scholarship at APC. Andrew May acknowledges the support of an STFC PhD Studentship.
\end{acknowledgements}


\end{document}

%% file: Authors.tex
\author{M. Piat	$^{1}$, B. B\'elier $^{2}$, L. Berg\'e	$^{3}$, N. Bleurvacq	$^{1}$, C. Chapron	$^{1}$, S. Dheilly	$^{1}$, L. Dumoulin	$^{3}$, M. Gonz\'alez	$^{4}$, L. Grandsire	$^{1}$, J.-Ch. Hamilton	$^{1}$
% ...
}

\institute{
$^{1}$ Astroparticule et Cosmologie (CNRS-IN2P3), Paris, France;
$^{2}$ Centre de nanosciences et de nanotechnologies, Orsay, France;
$^{3}$ Centre de Spectrom\'etrie Nucl\'eaire et de Spectrom\'etrie de Masse (CNRS-IN2P3), Orsay, France;
$^{4}$ Ctr. At\'omico Bariloche e Instituto Balseiro, CNEA, Argentina;
$^{5}$ Laboratoire de l'Acc\'el\'erateur Lin\'eaire (CNRS-IN2P3), Orsay, France;
$^{6}$ Institut de Recherche en Astrophysique et Plan\'etologie (CNRS-INSU), Toulouse, France;
$^{7}$ Kavli Institute for Particle Astrophysics and Cosmology, Stanford, California, USA;
$^{8}$ Universit\`a di Roma La Sapienza, Roma, Italy; 
$^{9}$ Istituto Nazionale di Fisica Nucleare Roma 1 Section, Roma, Italy;
$^{10}$ Cardiff University, Cardiff, UK;
$^{11}$ Univ. Nacional de la Plata, Argentina;
$^{12}$ Instituto de Tecnolog\'ias en Detecci\'on y Astropart\'iculas, Argentina;
$^{13}$ University of Manchester, Manchester, UK;
$^{14}$ Universit\`a degli Studi di Milano Bicocca, Milano, Italy;
$^{15}$ Istituto Nazionale di Fisica Nucleare Milano Bicocca section, Milano, Italy;
$^{16}$ National University of Ireland, Maynooth, Ireland;
$^{17}$ University of Milan, Dept. of Physics, Milano, Italy;
$^{18}$ Comisi\'on Nacional De Energia At\'omica, Argentina;
$^{19}$ Richmond University, Richmond, VA, USA;
$^{20}$ Universit\`a di Roma Tor Vergata, Roma, Italy;
$^{21}$ Universidad Nacional de San Martin, San Martin, Argentina;
$^{22}$ Institut d'Astrophysique Spatiale (CNRS-INSU), Orsay, France;
$^{23}$ Agenzia Spaziale Italiana, Rome, Italy;
$^{24}$ Pontificia Universidad Catolica de Chile, Santiago, Chile;
$^{25}$ Instituto Argentino de Radioastronomi­a, Argentina;
$^{26}$ California Institute of Technology, Pasadena, California, USA;
$^{27}$ Istituto Nazionale di Fisica Nucleare Pisa Section, Pisa, Italy;
$^{28}$ University of Wisconsin, Madison, WI, USA;
$^{29}$ Brown University, Providence, RI, USA;
$^{30}$ Regional Noroeste (CNEA), Argentina. 
\email{piat@apc.univ-paris7.fr}}

\authorrunning{Piat et al} 

%% file: Intro.tex
The measurement of the curl component of the polarization of the Cosmic Microwave Background (CMB), the so called \emph{B-modes}, is one of the most difficult challenge of the Observationnal Cosmology. Its detection would provide valuable information about the primordial universe, inflation theory as well as quantum gravity. 
Currently observing or planned CMB experiments share the same detection strategy: they are imagers able to perform high sensitivity observations of the sky thanks to background-limited broadband detectors but are all subject to similar instrumental systematic effects.
In interferometers what is formed on the focal plane are the Fourier modes of the sky radiation produced by an array of spatially correlated antennas. Generally, such instrumental configurations suffer low sensitivity and the impossibility of exploiting background-limited broadband direct detectors. 
The great advantage of interferometric instruments resides in their capability of controlling systematic effects. Measurements of the scalar component of the CMB polarization, the E-modes, have been performed in the past with such instrumental configuration (CBI [\cite{Padin02}], DASI [\cite{Halverson98}]).

In this context, the \emph{Q and U Bolometric Interferometer for Cosmology} (QUBIC) currently presents a unique working strategy, combining together the sensitivity of bolometric detectors together with the control of systematic effects typical of interferometers [\cite{Battistelli_LTD18}]. 



This paper focus on the description and characterization of the QUBIC detection chain. 
In section \ref{sec:instru}, we present the architecture of the instrument and its basic principle of operation while section \ref{sec:detchain}  describe the detection chain in more details. The recent results, which demonstrated the operation of bolometric interferometry, are presented in section \ref{sec:charact}.

%% file: Instru.tex
\begin{figure}[htbp]
\begin{center}
\includegraphics[width=1\linewidth, keepaspectratio]{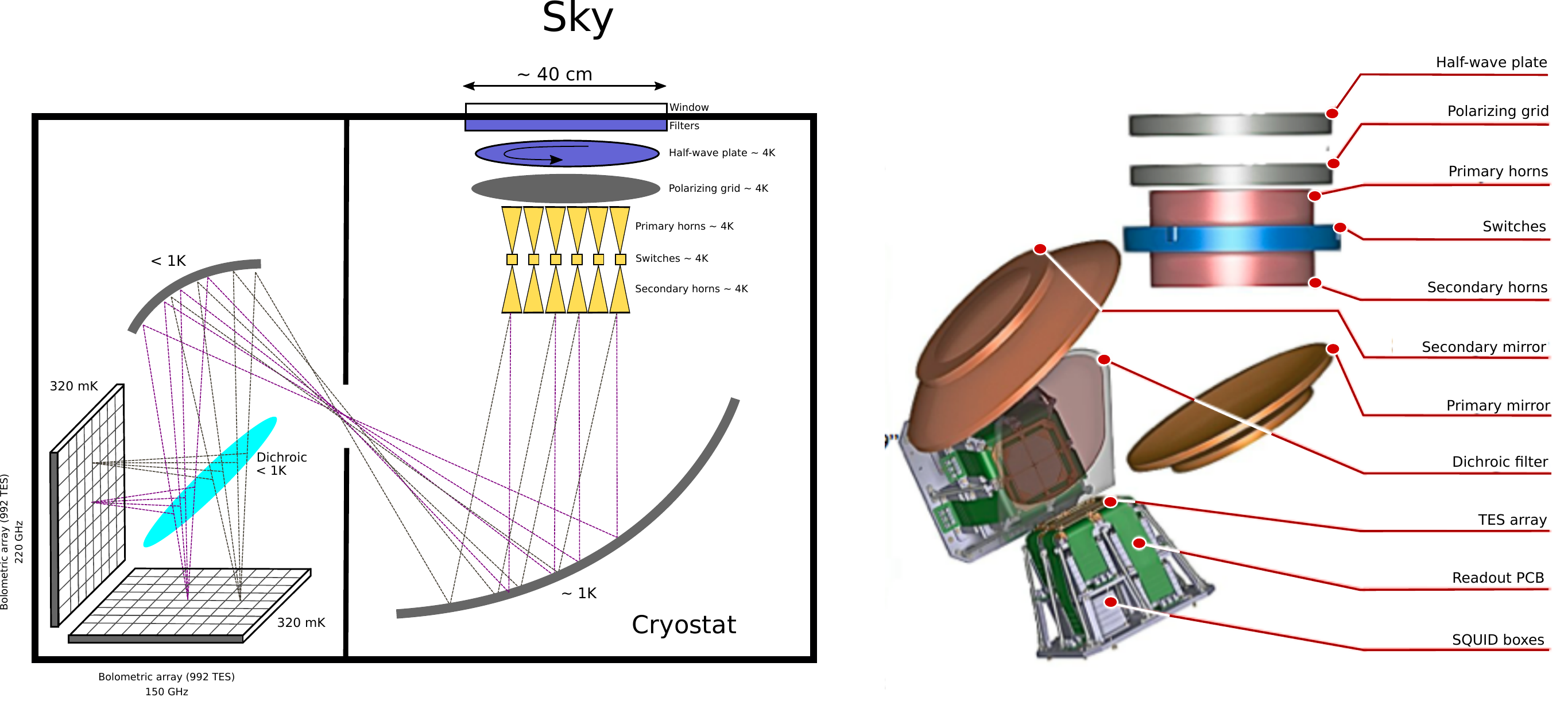}
\caption{{\it Left}: Schematic of the
      {\it QUBIC} instrument. The window aperture is about 40\,cm; the
      cryostat is about 1.41\,m in diameter and 1.51\,m in height;
      {\it Right}: 3D rendering of the inner part of the cryostat. TES,
      Transition Edge Sensor. (Color figure online.) 
      \label{fig_qubic_schematic}}
\end{center}
\end{figure}

QUBIC is an international ground based experiment dedicated to the observation of CMB polarisation. It will be deployed in Argentina, at the Alto Chorrillos mountain site (altitude of 4869 m a.s.l.) near San Antonio de los Cobres, in the Salta province. 
A full description of the QUBIC instrument is given in these proceedings [\cite{Battistelli_LTD18}] and in [\cite{Daniele2019, Creidhe2018, Andy2018, Andrea2016, Piat2012, Ellia2011}]. 
\\
A schematic of QUBIC is shown in Fig.~\ref{fig_qubic_schematic}. 
The signal from the sky enters the cryostat through a UHMW-PE  (Ultra-High Molecular Weight Polyethylene)  window [\cite{Dalessandro2018}].  Then, a rotating Half-Wave Plate (HWP)
modulates the polarization, and a polarizing grid selects one of the
two linear polarization components. An array of 400 back-to-back
corrugated horns collects the radiation and re-images it onto a
dual-mirror optical combiner that focuses the signal onto two
orthogonal TES detectors focal planes. A dichroic filter placed
between the optical combiner and the focal planes selects the two
frequency bands, centered at 150 GHz and 220 GHz. 
The detection chain is described in details in section \ref{sec:detchain}.
\\
In QUBIC, the optical combiner focuses the radiation emitted by the secondary horns onto the two focal planes so that the image that forms on the detector arrays is the result of the interference arising from the sum of the fields radiated from each of the 400 apertures.
In this sense a bolometric interferometer can be considered a synthesized imager: the image on the focal plane is the sky signal in intensity and polarization (modulated by the HWP orientation) convolved with the so-called {synthetic beam}.
\\
\\
QUBIC has two configurations: the ``Technological Demonstrator'' (TD)  and the ``Full Instrument'' (FI). The TD and FI share the same cryostat and cryogenics but the TD has only one-quarter of the 150-GHz TES focal plane (256 TESs), an array of 64 horns and switches and a smaller optical combiner.  The QUBIC TD is used to demonstrate the feasibility of the bolometric interferometry. 
%
%
It has been extensively tested in the last months at the APC laboratory in Paris. The instrument in now being upgraded with an improved sub-K cryogenic system and toward the FI hardware.
QUBIC will be shipped to Argentina and installed at the site for a first-light  foreseen at the beginning of 2020. 

%% file: Det_chain.tex
\begin{figure}[htbp]
\begin{center}
\includegraphics[width=0.54\linewidth, keepaspectratio]{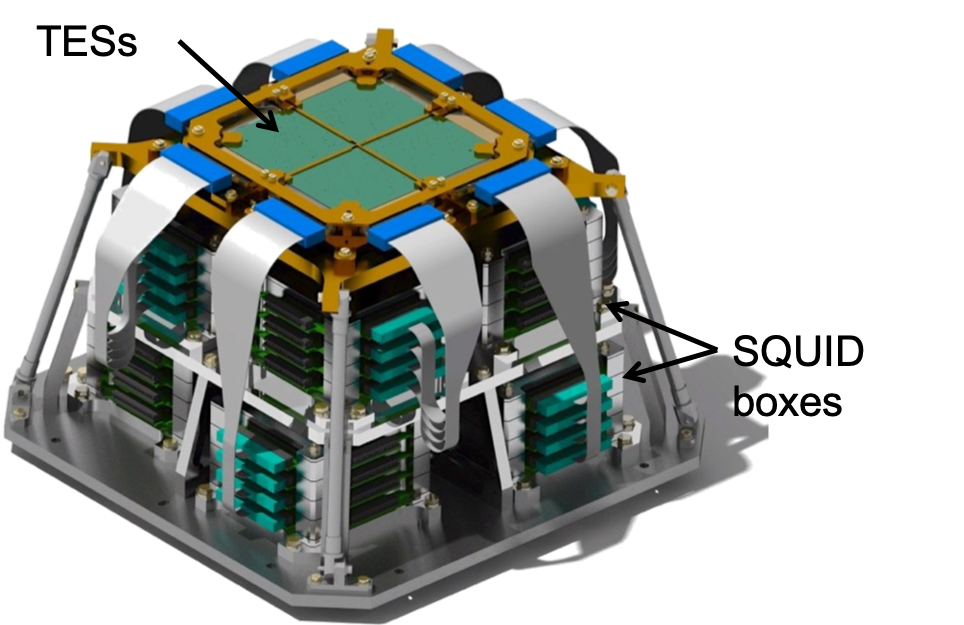}
\includegraphics[width=0.45\linewidth, keepaspectratio]{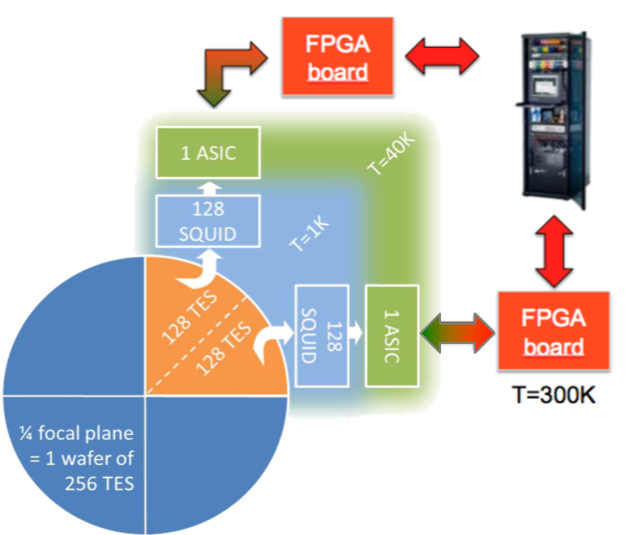}
\caption{{\it Left}: QUBIC cryo-mechanical structure which supports one TES focal plane at 350mK on top and the SQUID boxes at 1K below. {\it Right}: Architecture of the QUBIC detection chain for one focal plane of 1024 channels. (Color figure online.)
\label{Archi_det_chain}}
\end{center}
\end{figure}

The QUBIC detection chain architecture is shown on Fig.~\ref{Archi_det_chain}. Each focal plane is composed of four 256-pixel arrays
assembled together to obtain 1024-pixel detector. For each quarter focal plane, two blocks of 128 SQUIDs are used at 1K in a 128:1 Time Domain Multiplexing (TDM) scheme. Each block is controlled and amplified by an ASIC cooled to 40K while a warm FPGA board ensure the control and acquisition of the signal to the acquisition computer. 
%
\subsection{TES}
\begin{figure}[htbp]
\begin{center}
\includegraphics[width=4.9cm, keepaspectratio]{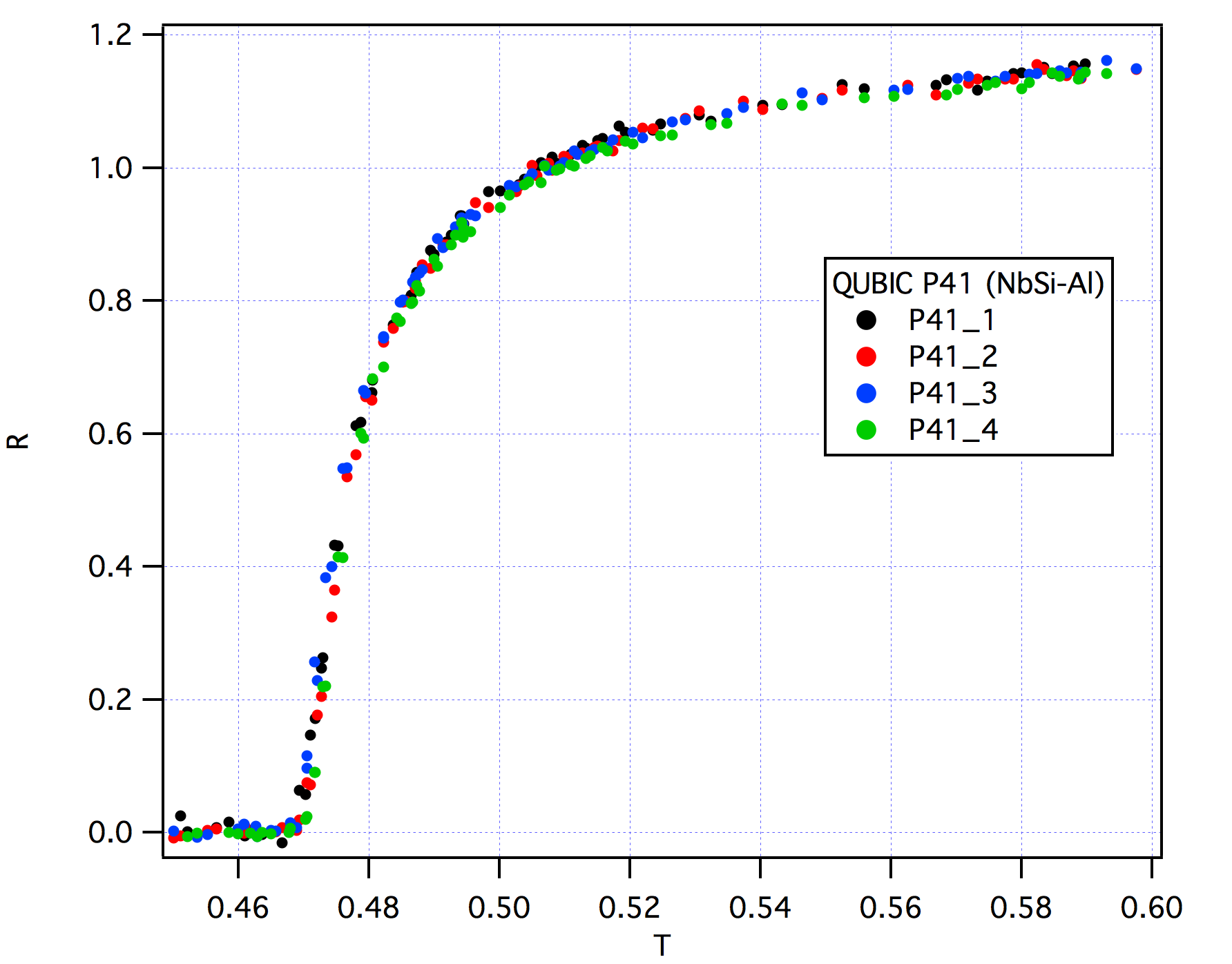}
\includegraphics[width=4.6cm, keepaspectratio]{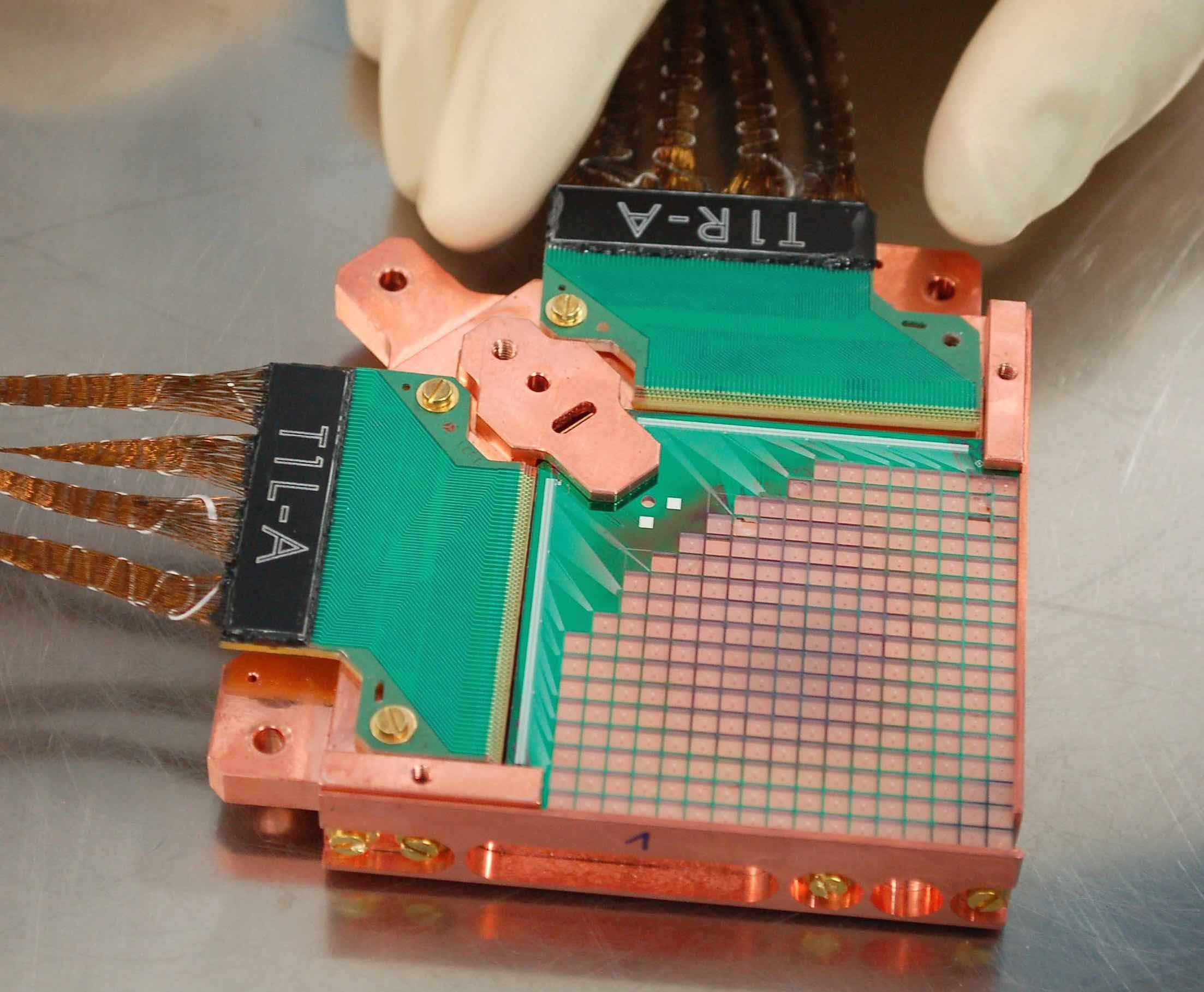} 
\caption{{\it Left}: Superconducting transition characteristics of four Nb\textsubscript{0.15}Si\textsubscript{0.85} TESs distributed far away from each other on a 256 pixel array. {\it Right}: Pictures of the 256 TES array  being integrated. (Color figure online.)
\label{testransi}}
\end{center}
\end{figure}

The detectors are  Transition Edge Sensors (TES) made with a Nb\textsubscript{x}Si\textsubscript{1-x} 
amorphous thin film (x${\approx}$0.15 in our case). 
Its transition temperature T\textsubscript{c} of about 500 mK (Fig.~\ref{testransi}) can be adapted by changing the composition $x$ of the compound. The normal state resistance R\textsubscript{n} is adjusted to about 1$\Omega$ with interleaved electrods for optimum performances. To adapt to the optics, the pixels have 3 mm spacing while the membranes structure is 2.7 mm wide without any sensitivity to polarization.
%
The low thermal coupling between the TES and the cryostat is obtained using 500 nm thin SiN suspended membranes, which exhibit thermal conductivities in the range 50-500 pW/K depending on the precise pixel geometry and temperature. The Noise Equivalent Power (NEP) is of the order of $5.10^{-17}W/\sqrt{Hz}$ at 150GHz with a natural time constant of about 100ms [\cite{salatino}]. 
Light absorption is achieved using a Palladium metallic grid placed in a quarter wave
cavity in order to optimize the  absorption efficiency. The backshort distance of 400 ${\mu}$m has been chosen after electro-magnetic simulations in order to have absorption higher than 94\% at both 150 and 220GHz.
%
\\
The 256-pixel array is finally integrated within the focal plane holder and electrically connected to a Aluminium Printed Circuit Board (PCB, provided by Omni Circuit Boards\footnote{www.omnicircuitboards.com}) using ultrasonic bonding of Aluminium wires (Fig.~\ref{testransi}). The detailed fabrication process of the QUBIC detectors is given in these proceedings [\cite{Marnieros_LTD18}]. The latest upgrade of the production process allows excellent fabrication quality with a dead pixels yield as low as 5\%. 
%
\subsection{SQUIDs}
\begin{figure}[htbp]
\begin{center}
\includegraphics[width=7cm]{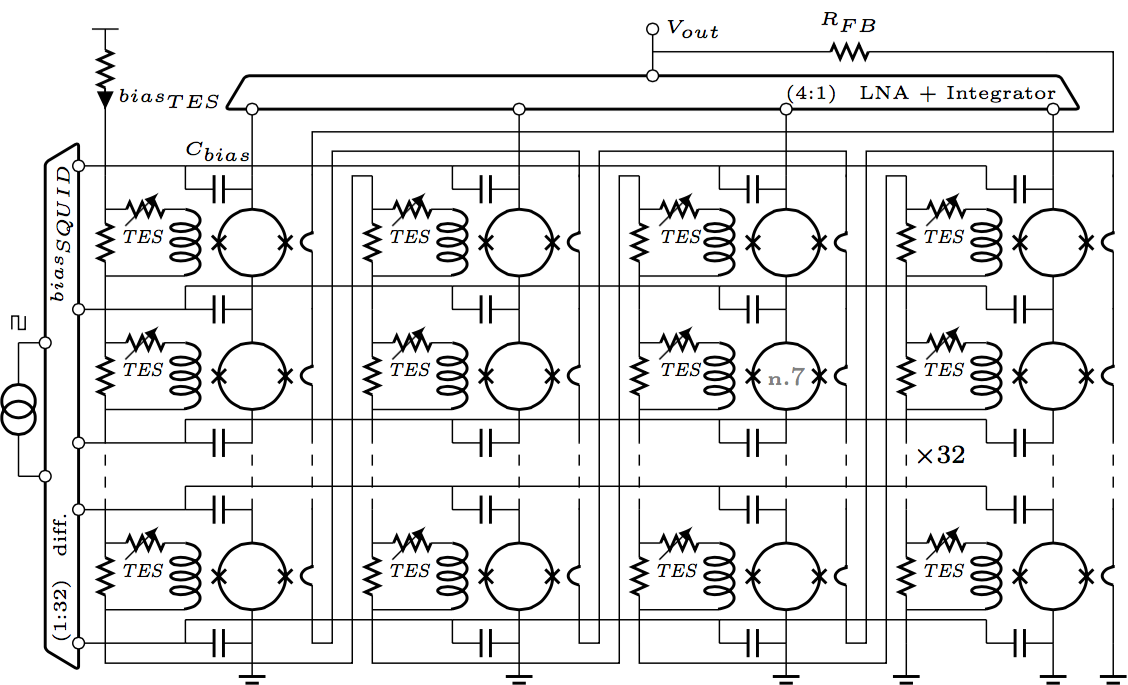} 
\includegraphics[width=5cm]{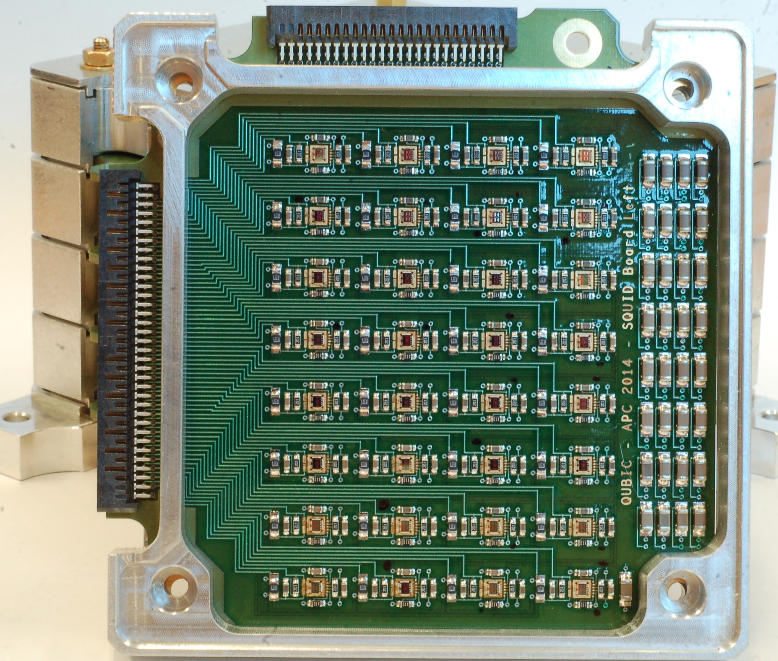} 
\caption{{\it Left}: Topology of the
128 to 1 multiplexer sub-system (4$\times$32 SQUIDs + 1 ASIC). {\it Right}: Integration of 32 SQUIDs (1 column) with bias capacitors and filter devices. (Color figure online.)
\label{fig84}}
\end{center}
\end{figure}

The SQUID multiplexer is composed of 4 columns of 32
SQUIDs biased with capacitors in order to reduce power dissipation and noise (Fig.~\ref{fig84}). 
The SQUIDs are based on a slightly modified SQ600S commercial design provided by StarCryoelectronics\footnote{starcryo.com} in order to reduce the area of each dies.  
Visual inspections and room temperature tests with a probe-station are used to select the SQUIDs before integration on a specific PCB. One SQUID PCB is composed of 32 SQUIDs and is integrated in an aluminium box. The architecture therefore uses 4 of this PCB boxes to readout 128 pixels. As shown in Fig. \ref{Archi_det_chain}, a stack of 8 SQUID boxes is installed at 1K below the TESs in the cryo-mechanical structure, surrounded with a Cryophy\footnote{www.aperam.com} magnetic shield.
%
%
%
\subsection{{ASIC}}
\label{sect_asic}
The ASIC is designed in full-custom using CADENCE CAD tools. 
The used technology is a standard $0.35\mu$  BiCMOS SiGe from Austria MicroSystem (AMS).  
This technology consists of p-substrate, 4-metal and 3.3~V process. It includes standards complementary MOS transistors
and high speed vertical SiGe NPN Heterojunction Bipolar Transistors (HBT). Bipolar transistors are preferentially used
for the design of analog parts because of their good performances at cryogenic temperature [\cite{Prele2018}]. 
The design of the ASIC is based on pre-experimental characterizations results, and its performance at cryogenic temperature is extrapolated from simulation results obtained at room temperature, using CAD tools.
\begin{figure}[htbp]
\begin{center}
\includegraphics[width=0.9\linewidth, keepaspectratio]{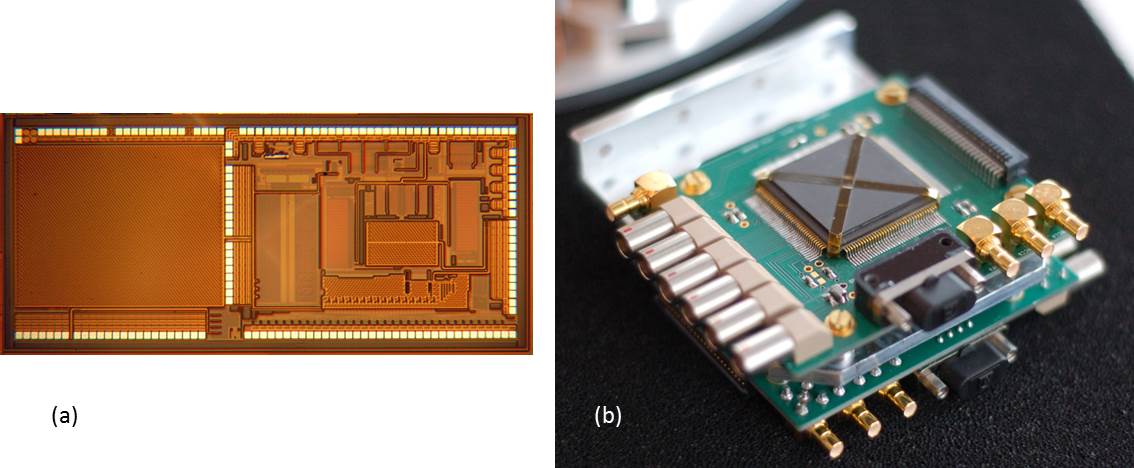} 
\caption{{\it Left}:  Microphotography of cryogenic ASIC
designed to readout 4$\times$32 TES/SQUID pixels. {\it Right}:  ASIC module assembly used for QUBIC experiment. (Color figure online.)
\label{fig:cryoasic}}
\end{center}
\end{figure}
%
Each ASIC board for QUBIC (shown on Fig.~\ref{fig:cryoasic}) has a power dissipation of typically 16 mW and is placed on the 40K stage. The ASIC integrates all parts needed to achieve the readout, the multiplexing and the control of an array of up to 128 TESs/SQUIDs. 
It includes a differential switching current source to address sequentially 32 lines of SQUIDs,
achieving a first level of multiplexing of 32:1. In this configuration, the SQUID are AC biased through capacitors which allows a good isolation (low crosstalk between SQUID columns) and no power dissipation. 
A cryogenic SiGe low noise amplifier ($e_n=$0.3 nV/$\sqrt{Hz}$, gain=70, bandwidth of about 6MHz in simuations) with 4 multiplexed inputs, performs a second multiplexing stage between each column. 

This cryogenic ASIC includes also the digital synchronization circuit of the overall
multiplexing switching (AC current sources and multiplexed low noise amplifier). A serial protocol allows to focus on sub-array as well as to adjust the amplifiers and current sources with a reduced
number of control wires. 
As the digital side takes a large part, we have developed a full custom CMOS digital library dedicated to cryogenic application and ionizing environments (rad-hard full custom digital library) [\cite{Prele2018}]. 
%
\subsection{{Warm electronics and acquisition software}}
%
The warm electronics is based on FPGA boards called NetQuiC, one
for each ASIC. These boards are connected to the acquisition computer via a network switch. 
Each NetQuiC board is based on a differential amplifier (gain=100, bandwidth limited to 1MHz with a second order low-pass filter), a 2MHz 16 bits ADC, 7 16 bits DACs and a Xilinx Spartan 6 FPGA (XEM6010 board from Opal Kelly).
The FPGA firmware programmed in VHDL takes in charge the following tasks: ASICs control, management of the TCP/IP connection with the acquisition computer, acquisition of scientific signal with the ADC, bias generation and digital Flux Locked Loop (FLL) control.
%
%
The acquisition software called QUBIC Studio is the single interface to deal with the readout, the control command software and the data storage. Its core is  the generic tool called ``dispatcher''  developed at IRAP that is a real-time-oriented acquisition system widely used on various experiments such as Solar Orbiter, SVOM/ECLAIRS and PILOT. 

%% file: Charact.tex
The QUBIC TD test campaign has been done first in blind configuration (the 40K filters being closed) in December 2018 and afterward with all filters open from mid-January to beginning of July 2019. Unfortunately, the 1K fridge was out of order during the second run which imply a 1K stage temperature of only 4K. The detector stage therefore reached only 370mK, not low enough to be in the superconducting state for the detectors. The TESs were nevertheless responsive which allow us to make a first complete optical characterization of the instrument (see [\cite{Battistelli_LTD18}] for further details). We report here the characterizations made on the detection chain.
\subsection{ASICs}
Both ASICs 1 and 2 have been functionally tested and characterized at low temperature during the TD test campaigns. Low noise multiplexed amplifier characterizations have been investigated using a vector analyzer. 
A white noise level of 0.3 nV/$\sqrt{Hz}$ and a knee frequency of about 400Hz were measured at 70K with a differential voltage gain of 70 (measured in a specific cryogenic test bench). 
%
%
\subsection{SQUIDs}
\begin{figure}[htbp]
\begin{center}
\includegraphics[width=0.47\linewidth, keepaspectratio]{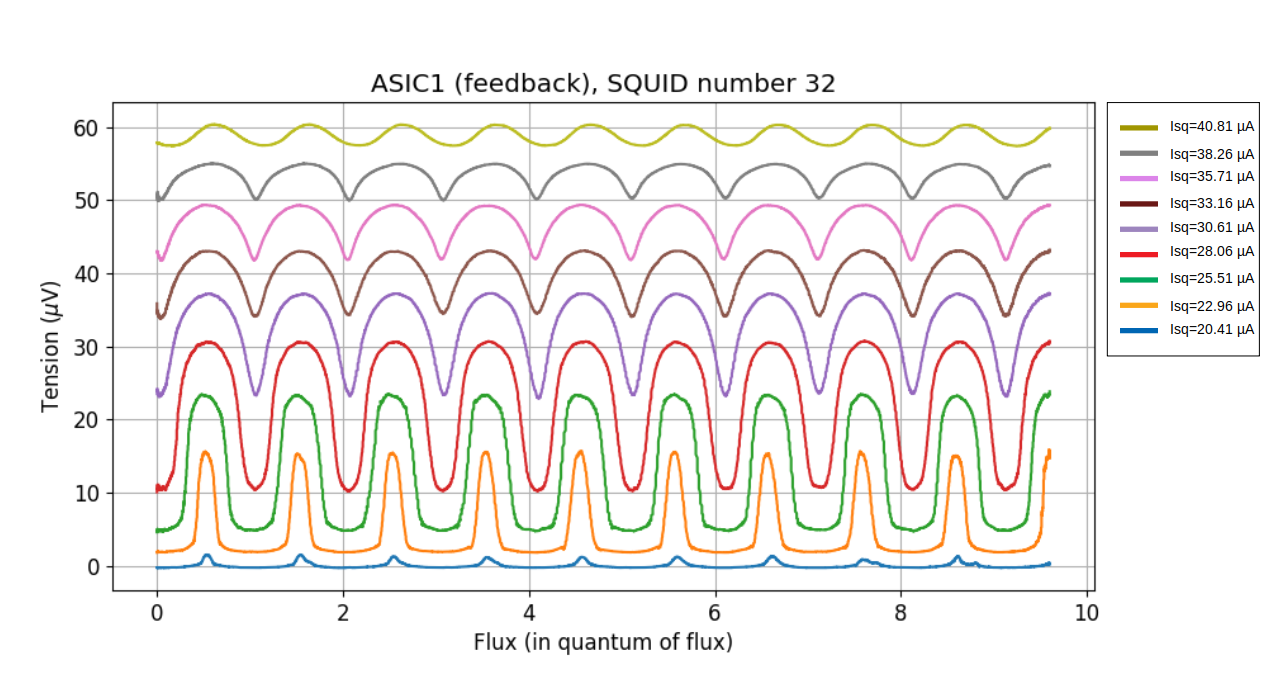} 
\includegraphics[width=0.5\linewidth, keepaspectratio]{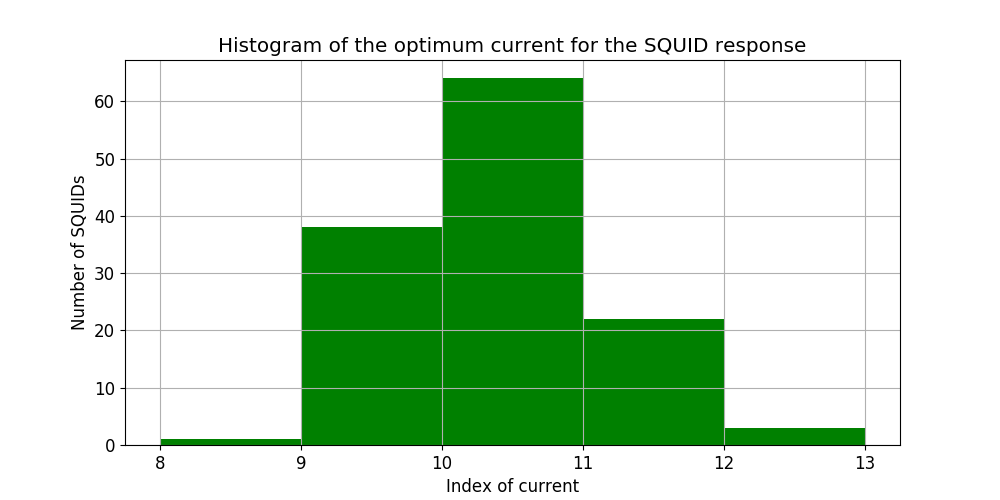} 
\caption{\textit{Left}: typical flux response of a QUBIC SQUID for different currents. \textit{Right}: histogram of the SQUID response as a function of current (expressed in digital unit). (Color figure online.)
\label{fig:SQUIDs_meas}}
\end{center}
\end{figure}

The SQUID bias current source in the ASIC is coded on 4 bits and so we have 16 possible values spanning the range (0$\div$40.8)$\mu$A. The feedback coil being biased with a sine wave signal produced by the FPGA, this allows to measure both the the V-I and the V-$\Phi$ curves (see Fig.~\ref{fig:SQUIDs_meas}).
The optimal bias current is selected as the value which maximizes the SQUIDs response for the greatest number of SQUIDs read out with the same ASIC. As expected, the optimal bias current is close to the critical current value (26\,$\mu$A for our SQ600S SQUIDs). The final choice in bias current allows to readout more than 86\% of the SQUIDs with a voltage swing higher than 10$\mu$V.
\subsection{TESs}
\begin{figure}[htbp]
\begin{center}
\includegraphics[width=0.44\linewidth, keepaspectratio]{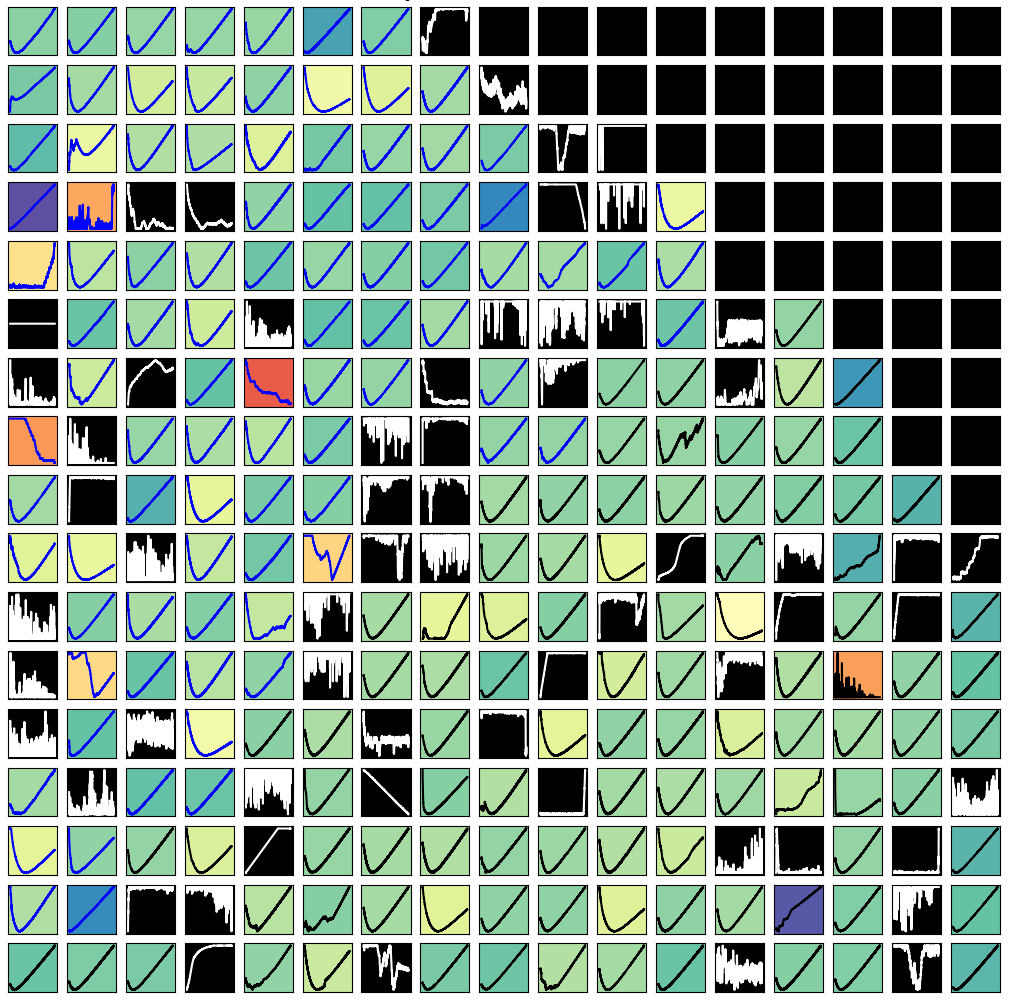} 
\includegraphics[width=0.455\linewidth, keepaspectratio]{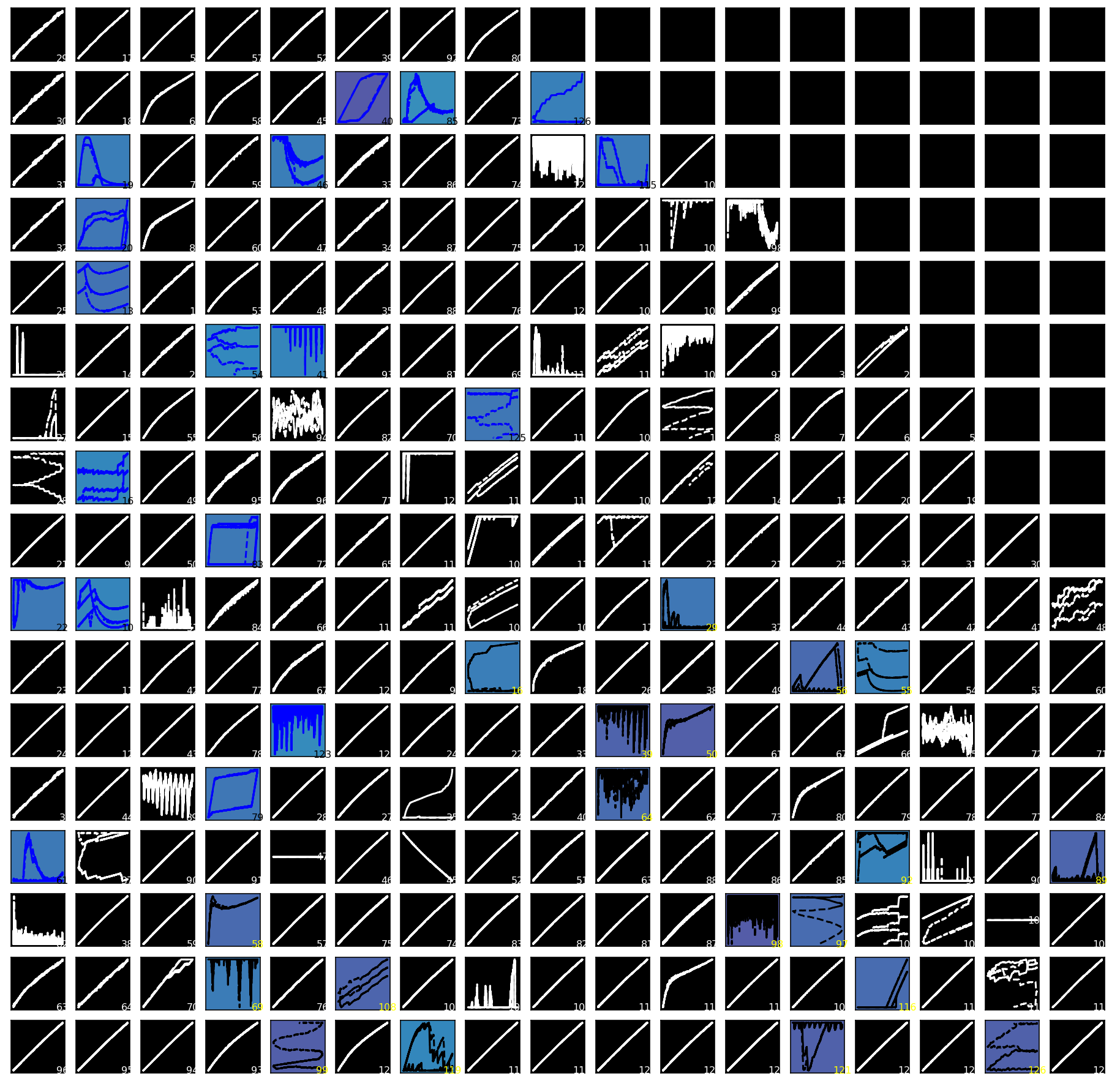} 
\caption{{\it Left:} I-V curves of the QUBIC P87 TES array measured in the blind configuration at 348mK. The color code indicate the turn-over bias voltage. {\it Right:} same array measured in open configuration at 382mK. (Color figure online.)
\label{fig:TESmeas}}
\end{center}
\end{figure}

The TESs have been characterized both in the blind and open configuration (Fig.~\ref{fig:TESmeas}). In the later case, the power background and the limited temperature of the detector stage did not allow to reach the superconducting state. 
In the closed configuration, it demonstrate the operation of the TES in ETF mode in the lowest bias voltage as expected. 
The dynamic thermal conductance is measured at about $250\;$pW/K leading to a NEP of $5-6$.$10^{-17}$W/$\sqrt{Hz}$ at 350mK.
Moreover, a yield of about 74\% is obtained on this array. In the open configuration, the fact that the R(T) curve still has a slope in the higher part as seen in Fig.~\ref{testransi} allow us to get a high enough response to do a  full optical characterization with a calibration source\cite{Battistelli_LTD18}  without the nominal sensitivity.

%

%% file: Ccl.tex
The QUBIC detection chain is based on two 1024 NbSi TES field arrays and an original 128:1 TDM readout scheme using a cold SiGe ASIC. This detection chain was functionally validated during the QUBIC TD calibration campaigns at APC. Although a cryogenic problem did not allow to cool the 1K stage at its nominal temperature, a first complete optical characterizations of this new instrument has been performed.
These measurements open the path toward the self-calibration technique and the spectro-imaging capabilities of QUBIC. 
The instrument is now being upgraded toward its final configurations, with an expected first light on site in 2020.